\documentstyle[aps,preprint]{revtex}

\begin{document}
\draft
\tighten
\title
{\bf Superconductivity in a Toy Model of the Pseudogap State}
\author{E.Z.Kuchinskii,\ M.V.Sadovskii}
\address
{Institute for Electrophysics,\\ Russian Academy of Sciences,\ 
Ural Branch,\\ Ekaterinburg,\ 620049,\ Russia\\
E-mail:\ kuchinsk@ief.uran.ru,\ sadovski@ief.uran.ru} 
\maketitle


\begin{abstract}
We analyze superconducting state (both $s$ and $d$ -- wave) in a simple
exactly solvable model of pseudogap state, induced by short -- range order
fluctuations (e.g. antiferromagnetic), which is based upon model Fermi --
surface with ``hot patches''. It is shown that superconducting energy gap
averaged over these fluctuations is non -- zero even for the temperatures
larger than mean -- field $T_c$ of superconducting transition in a sample as
a whole. For temperatures $T>T_c$ superconductivity apparently exists within
separate regions (``drops''). We study the spectral density and the density 
of states and demonstrate that superconductivity signals itself in these
already for $T>T_c$, while at $T_c$ itself nothing special happens from this
point of view. These anomalies are in qualitative agreement with a number
experiments on underdoped cuprates.
\end{abstract} 

\newpage
\narrowtext
\section{Introduction}

Among different anomalies of electronic properties of high -- temperature
superconductors especially interesting are those of the pseudogap state 
observed, mainly, in the underdoped region of the phase diagram \cite{Tim,Ran}.  
Pseudogap anomalies are seen in a number of experiments, such as optical
conductivity, NMR, inelastic neutron scattering, angle -- resolved 
photoemission (ARPES) etc. \cite{Tim}. Especially striking evidence of the
existence of this state is observed in ARPES -- experiments \cite{Tim,RC},
which demonstrate essentially anisotropic changes of the spectral density of
current carriers in a wide temperature region both in normal and
superconducting state of these systems. Maximum of these anomalies is
observed around ($\pi,0$) point of the Brillouin zone, while in the
direction of its diagonal (close to ($\pi,\pi$) -- point) these are
practically absent. Qualitatively speaking these anomalies signify the
complete ``destruction'' of the Fermi surface in the vicinity of ($\pi,0$) --
point, while Fermi liquid behavior is conserved in the direction of diagonal.
In this sense it is usually stated that the pseudogap possesses
``$d$ -- wave'' symmetry, similar to that of the superconducting gap in these
compounds \cite{Tim,Ran,RC}. At the same time, the fact that pseudogap
anomalies are observed up to temperatures $T\sim T^*$ which are significantly
larger than superconducting $T_c$, can be an evidence for quite different
nature of these anomalies, not connected to superconducting pairing. 
This conclusion may be further supported by the fact that pseudogap state is
observed mainly in underdoped cuprates, i.e. for compositions which are close
to antiferromagnetic phase.

There are two major approaches to theoretical understanding of the pseudogap
state of high -- temperature superconductors. One is based upon rather
popular idea of Cooper pair formation above the temperature of
superconducting transition \cite{Ran,Gesh,EK,Lev,Lok}. The other assumes that
the pseudogap state is somehow induced by fluctuations of antiferromagnetic
short -- range order (see e.g. \cite{Kam,Benn,Dei,Sch,KS}).

The majority of theoretical works are devoted to the study of pseudogap state
in the normal phase of cuprates at $T>T_c$. In a recent paper \cite{PS} 
a very simple exactly solvable model of the pseudogap state was proposed,
based upon the picture of ``hot'' (nesting) patches on the Fermi surface.
Within this model a derivation of Ginzburg -- Landau expansion was given for
different types of Cooper pairing with qualitative analysis of pseudogap
effects (induced by AFM short -- range order fluctuations)
on the main superconducting properties close to $T_c$. The present work
extends this model to the study of anomalies of superconducting state
within the pseudogap for all temperatures $T<T_c$.
 
\section{Model of the pseudogap state.}

We shall consider a greatly simplified model of the pseudogap state
\cite{PS}, which is based on the idea of well -- developed fluctuations
of antiferromagnetic (AFM, SDW) short -- range order
\footnote{Note that our analysis can also be applied to the case of
CDW short -- range order and other similar models.}, 
which is qualitatively similar to the ``hot spots'' model of Ref. \cite{Sch}. 
We assume that the Fermi surface of two -- dimensional system of electrons
has the form shown in Fig. 1. Similar Fermi surface was in fact observed
in a number of ARPES experiments on cuprate superconductors (e.g. see quite 
recent papers \cite{Onell,Shen}.). Fluctuations of short -- range order are
assumed to be static and Gaussian with the following correlation function:  
(cf. \cite{Kam}):  
\begin{equation} 
S({\bf q})=\frac{1}{\pi^2}\frac{\xi^{-1}}{(q_x-Q_x)^2+\xi^{-2}} 
\frac{\xi^{-1}}{(q_y-Q_y)^2+\xi^{-2}}
\label{fluct}
\end{equation}
where $\xi$-- correlation length of these fluctuations, and the scattering
vector is taken to be  either
$Q_x=\pm 2k_F$,\ $Q_y=0$ or $Q_y=\pm 2k_F$,\ $Q_x=0$.  
We also assume that these fluctuations interact only with electrons from
the flat (``hot'') patches on the Fermi surface shown in Fig. 1, and this
scattering is in fact of one -- dimensional nature.
The effective interaction with fluctuations will be described by the value of
$(2\pi)^2W^2S({\bf q})$, where parameter $W$ of dimensions of energy is
defining the energy scale (width) of the pseudogap
\footnote{We can say that we introduce an effective electron -- fluctuations
coupling constant of the form: 
$W_{\bf p}=W[\theta(p_x^0-p_x)\theta(p_x^0+p_x)+\theta(p_y^0-p_y) 
\theta(p_y^0+p_y)]$.}.  
The choice of the scattering vector ${\bf Q}=(\pm 
2k_F,0)$ or ${\bf Q}=(0,\pm 2k_F)$ implicitly assumes that fluctuations
are incommensurate (generalization to the commensurate case is also
possible \cite{PS}, but we shall not discuss it here).  

In the limit of $\xi\to\infty$ this model allows an exact solution by
methods proposed (for one -- dimensional case) in Ref. \cite{C74}. 
For the case of finite $\xi$ a ``nearly'' exact solution may be constructed
\cite{Sch,KS}, directly generalizing a one -- dimensional Ansatz of
Refs. \cite{C79,C91}. In this work we consider only maximally simplified
variant of this model with $\xi\to\infty$, when effective interaction with
fluctuations (\ref{fluct}) takes the simplest possible form
\footnote{Let us stress that due to the Gaussian nature of fluctuations
the limit of $\xi\to\infty$ does not assume any kind of long -- range order
in the system.}:  
\begin{equation} 
(2\pi)^2W^2\left\{\delta(q_x\pm 2p_F)\delta(q_y)
+\delta(q_y\pm 2p_F)\delta(q_x)\right\}
\label{WW}
\end{equation}
In this case we can easily sum all diagrams of the perturbation series for an
electron scattered by these fluctuations \cite{C74} and obtain one -- 
particle Green's function in the following form \cite{PS}:  
\begin{equation} 
G(\epsilon_n,p)=\int \limits_0^\infty dD {\cal P}(D)
\frac{i\epsilon_n+\xi_p}{(i\epsilon_n)^2-\xi_p^2-D(\phi)^2}, 
\label{fgrina} 
\end{equation} 
where $\xi_p=v_F(|{\bf p}|-p_F)$ ($v_F$ - Fermi velocity), 
$\epsilon_n=(2n+1)\pi T$, and fluctuating dielectric gap $D(\phi)$
is different from zero only on the ``hot'' patches:
\begin{equation} 
D(\phi)=\left\{ \begin{array}{ll} D & 
,0\leq\phi\leq\alpha,\>\frac{\pi}{2}-\alpha\leq\phi\leq\frac{\pi}{2} \\ 0 & 
,\alpha\leq\phi\leq\frac{\pi}{2}-\alpha \end{array} \right.  
\label{D} 
\end{equation}
where $\alpha=arctg(\frac{p_y^0}{p_F})$, $\phi$ -- is polar angle, defining 
the direction of vector ${\bf p}$ in ($p_x,p_y$) -- plane. For other values
of $\phi$  the value of $D(\phi)$ is defined similarly to (\ref{D})
by obvious symmetry considerations.

The amplitude of dielectric gap $D$ is random and distributed according to
Rayleigh \cite{C79} (its phase is also random and distributed homogeneously
on the interval $(0,2\pi)$):
\begin{equation}
{\cal P}(D)=\frac{2D}{W^2}\exp\left(-\frac{D^2}{W^2}\right)
\label{Rayl}
\end{equation}
Thus on the ``hot'' patches this Green's function has the form of ``normal''
Gor'kov's function, averaged over fluctuations of dielectric gap
$D$, distributed according to (\ref{Rayl}). ``Anomalous'' Gor'kov's functions
on these ``dielectrized'' patches are equal to zero (due to random phases
of dielectric gap $D$) in accordance with the absence of any long -- range
order. However, the average values of the pairs of these functions are non
zero and contribute to the two -- particle Green's function \cite{C74,PS}.
Varying $\alpha$ in (\ref{D}) within the interval 
$0\leq\alpha\leq\frac{\pi}{4}$, we actually change the size of ``hot'' 
patches on the Fermi surface where the nesting condition $\xi_{p-Q}=-\xi_p$
is fulfilled. In particular, the value of $\alpha=\pi/4$ corresponds to the
square -- like Fermi surface. Outside ``hot'' patches (second inequality
in (\ref{D})) the Green's function (\ref{fgrina}) just coincides with that of
free electrons.

Our results for electronic density of states and spectral density following
from (\ref{fgrina}) were given in Ref. \cite{PS} and demonstrated the
the pseudogap of characteristic width of $\sim 2W$ as well as non-Fermi
liquid behavior on the ``hot'' patches. In the case of finite correlation
lengths $\xi$ the Green's function on these patches is expressed via certain
continuous fraction \cite{C99} (cf. similar results in Refs. 
\cite{C79,C91,Sch,KS}) and spectral density demonstrates more ``smeared''
(in comparison with the case of $\xi\to\infty$) behavior with diminishing 
$\xi$, which is described in detail in Refs. \cite{C91,Sch,KS}. In Ref. 
\cite{C99} this model was applied to calculations of optical conductivity of
two -- dimensional system in the pseudogap state.

\section{Superconductivity in the Pseudogap State.}

Consider now the case for superconductivity in this model. Let us assume that
superconducting pairing is induced by an attractive interaction of the
following simplest form \cite{PS}:
\begin{equation}
V({\bf p,p'})=V(\phi,\phi')=-Ve(\phi)e(\phi'),
\label{VV}
\end{equation}
where $\phi$ -- is as above an angle defining the direction of electronic
momentum ${\bf p}$ in the plane, while for $e(\phi)$ we take the simplest
model dependence: 
\begin{equation} 
e(\phi)= \left\{ 
\begin{array}{ll}
1 & (\mbox{ $s$-wave pairing})\\ 
\sqrt{2}\cos(2\phi) & (\mbox{ $d$-wave pairing})
\end{array}.
\right.
\label{ephi}
\end{equation}
Coupling constant $V$, as usual, is taken to be non zero in some energy
interval of the order of $2\omega_c$ around the Fermi level
(where $\omega_c$ --  is characteristic frequency of the quanta responsible
for attraction). In this case superconducting energy gap has the following
form: 
\begin{equation}
\Delta({\bf p})\equiv \Delta(\phi)=\Delta e(\phi).
\label{DD}
\end{equation}
Let us first consider superconductivity in a system with {\em fixed}
dielectric gap $D$ on ``hot'' patches of the Fermi surface.
The question of superconductivity in a system with partial dielectrization
of electronic spectrum was studied in a number of works
(see e.g. \cite{Kop,Ginz}). The case very similar to that considered here it
was analyzed in a paper by Bilbro and McMillan   
\cite{Bilb}, from which we can immediately use some of the results for
rather simple generalization for the case under study.

In particular, for the case of  $s$-wave pairing, superconducting gap $\Delta$
equation is:  
\begin{equation} 
1=\lambda\int_{0}^{\omega_c}d\xi \left\{\tilde\alpha\frac
{th\frac{\sqrt{\xi^2+D^2+\Delta^2(D)}}{2T}}{\sqrt{\xi^2+D^2+
\Delta^2(D)}}+(1-\tilde\alpha)\frac{th\frac{\sqrt{\xi^2+\Delta^2(D)}}{2T}}
{\sqrt{\xi^2+\Delta^2(D)}}\right\}
\label{gapsw}
\end{equation}
where $\lambda=VN_0(0)$ -- is the dimensionless pairing coupling constant
($N_0(0)$ -- is the density of states of free electrons at the Fermi level), 
while parameter $\tilde\alpha=4\alpha/\pi$ defines a fraction of ``hot''
patches on the Fermi surface.

First term in Eq. (\ref{gapsw}) corresponds to the contribution of ``hot''
(dielectrized) patches, where the electronic spectrum is \cite{Bilb}: 
$E_p=\sqrt{\xi^2_p+D^2+\Delta^2}$, while the second term gives the
contribution of ``cold'' (metallic) patches, where the spectrum is the
usual BCS -- like: $E_p=\sqrt{\xi^2_p+\Delta^2}$. Equation (\ref{gapsw})
defines superconducting gap $\Delta(D)$ for the fixed value of dielectric gap
$D$, which is non zero at ``hot'' patches.

In case of $d$-wave pairing, analogous equation is:
\begin{equation}
1=\lambda\frac{4}{\pi}\int_{0}^{\omega_c}d\xi 
\left\{\int_{0}^{\alpha}d{\phi}e^2(\phi)\frac
{th\frac{\sqrt{\xi^2+D^2+\Delta^2(D)e^2(\phi)}}{2T}}{\sqrt{\xi^2+D^2+
\Delta^2(D)e^2(\phi)}}+\int_{\alpha}^{\pi/4}d{\phi}e^2({\phi})
\frac{th\frac{\sqrt{\xi^2+\Delta^2(D)e^2(\phi)}}{2T}}
{\sqrt{\xi^2+\Delta^2(D)e^2(\phi)}}\right\}
\label{gapdw}
\end{equation}
From these equations it can be seen that $\Delta(D)$ diminishes with the 
growth of $D$, while $\Delta(0)$ coincides with $\Delta_0$ in the absence
of dielectrization on flat patches which appears at temperature
$T=T_{c0}$, defined by the equation:  
\begin{equation} 
1=\lambda\int_{0}^{\omega_c}d\xi \frac{th\frac{\xi}{2T_{c0}}}{\xi}
\label{tc0}
\end{equation}
both for $s$-wave and $d$-wave pairing.

For $D\to\infty$ first terms in Eqs. (\ref{gapsw}), (\ref{gapdw}) tend to 
zero, so that equations for $\Delta_{\infty}=\Delta(D\to\infty)$ are: 
\begin{equation} 
1=\lambda\int_{0}^{\omega_c}d\xi
(1-\tilde\alpha)\frac{th\frac{\sqrt{\xi^2+\Delta^2_{\infty})}}{2T}}
{\sqrt{\xi^2+\Delta^2_{\infty}}}\quad \mbox{($s$-wave pairing)}
\label{gapswinf}
\end{equation}
\begin{equation}
1=\lambda\frac{4}{\pi}\int_{0}^{\omega_c}d\xi 
\int_{\alpha}^{\pi/4}d{\phi}e^2({\phi})
\frac{th\frac{\sqrt{\xi^2+\Delta^2_{\infty}e^2(\phi)}}{2T}}
{\sqrt{\xi^2+\Delta^2_{\infty}e^2(\phi)}}\quad \mbox{($d$-wave pairing)}
\label{gapdwinf}
\end{equation}
Eq. (\ref{gapswinf}) coincides with gap equation for $D=0$ with
``renormalized'' coupling constant $\tilde\lambda=\lambda(1-\tilde
\alpha)$, so that for the case of $s$-wave pairing:
\begin{equation}
\Delta_{\infty}=\Delta_0(\tilde\lambda=\lambda(1-\tilde\alpha))
\label{delsinf}
\end{equation}
and non zero superconducting gap for $D\to\infty$ appears at
$T<T_{c\infty}$:
\begin{equation}
T_{c\infty}=T_{c0}(\tilde\lambda=\lambda(1-\tilde\alpha)).
\label{tcsinf}
\end{equation}
In case of $d$-wave pairing from Eq.(\ref{gapdwinf}) we get:
\begin{equation}
T_{c\infty}=T_{c0}(\tilde\lambda=\lambda(1-\alpha_d))
\label{tcdinf}
\end{equation}
where
\begin{equation}
\alpha_d=\tilde\alpha+\frac{\sin\pi\tilde\alpha}{\pi}
\label{alphd}
\end{equation}
defines an ``effective'' fraction of flat patches in case of $d$-wave 
pairing. Thus, for $T<T_{c\infty}$ superconducting gap is non zero for
arbitrary values of $D$ and diminishes from $\Delta_0$ to $\Delta_{\infty}$ 
with the growth of $D$. For $T_{c\infty}<T<T_{c0}$ the gap is different from
zero only for $D<D_{max}$. Appropriate dependences of $\Delta$ on $D$ 
can be easily found by numerical solution of Eqs.(\ref{gapsw}) and 
(\ref{gapdw}).

In our model of the pseudogap state dielectric gap $D$ is not fixed but
random and is distributed according to (\ref{Rayl}). Accordingly the above
equations should be averaged over these fluctuations. We can, for example,
directly calculate the superconducting gap $<\Delta>$ averaged over the
fluctuations of $D$:
\begin{equation}
<\Delta> = \int_{0}^{\infty}dD{\cal P}(D)\Delta(D)=
\frac{2}{W^2}\int_{0}^{\infty}dDDe^{-\frac{D^2}{W^2}}\Delta(D)
\label{avergap}
\end{equation}
Here, dependences of $\Delta(D)$ described above immediately lead to the
conclusion that the averaged gap (\ref{avergap}) is in fact non zero up to
temperature $T=T_{c0}$, i.e. superconducting transition temperature in the
absence of any pseudogap anomalies. However, it is obvious that transition
temperature $T_c$ for a superconductor with pseudogap is lower than
$T_{c0}$ \cite{PS}. Thus, an apparently paradoxical behavior of $<\Delta>$ 
signifies, probably, the appearance in the system of local regions with
$\Delta\neq 0$ (superconducting ``drops'') induced by fluctuations of $D$ for
all temperatures $T_c<T<T_{c0}$, while coherent superconducting state appears
in the sample only for $T<T_c$. Of course the complete justification of
these qualitative picture can be obtained only in more realistic model with
finite correlation length $\xi$ of AFM -- fluctuations\footnote{Qualitatively
this situation resembles the appearance of inhomogeneous superconducting
state induced by strong fluctuations of the local density of states close to
the Anderson metal -- insulator transition \cite{BPS,C97}.}.
At the same time the simplicity of the current model with $\xi\to\infty$ 
allows immediately to obtain an exact result for $<\Delta>$.

To determine superconducting transition temperature $T_c$ in a sample as a
whole, we shall use the standard mean -- field approach (compare e.g. the
analogous approach for a superconductor with impurities \cite{C97}), assuming
the self -- averaging of the superconducting gap over fluctuations of $D$
(i.e. in fact independence of $\Delta$ on $D$ -- fluctuations). Then the
equations for mean -- field gap $\Delta_{mf}$ take the following form: 
\begin{equation}
1=\lambda\int_{0}^{\omega_c}d\xi \left\{\tilde\alpha\frac{2}{W^2}
\int_{0}^{\infty}dDDe^{-\frac{D^2}{W^2}}\frac
{th\frac{\sqrt{\xi^2+D^2+\Delta^2_{mf}}}{2T}}{\sqrt{\xi^2+D^2+
\Delta^2_{mf}}}+(1-\tilde\alpha)\frac{th\frac{\sqrt{\xi^2+\Delta^2_{mf}}}{2T}}
{\sqrt{\xi^2+\Delta^2_{mf}}}\right\} 
\label{gapswmf}
\end{equation}
for the case of $s$-wave pairing, and
\begin{eqnarray}
1=\lambda\frac{4}{\pi}\int_{0}^{\omega_c}d\xi\left\{ 
\int_{0}^{\infty}dDDe^{-\frac{D^2}{W^2}}
\int_{0}^{\alpha}d{\phi}e^2(\phi)\frac
{th\frac{\sqrt{\xi^2+D^2+\Delta^2_{mf}e^2(\phi)}}{2T}}{\sqrt{\xi^2+D^2+
\Delta^2_{mf}e^2(\phi)}}+\right.\nonumber\\
\left.+\int_{\alpha}^{\pi/4}d{\phi}e^2({\phi})
\frac{th\frac{\sqrt{\xi^2+\Delta^2_{mf}e^2(\phi)}}{2T}}
{\sqrt{\xi^2+\Delta^2_{mf}e^2(\phi)}}\right\}
\label{gapdwmf}
\end{eqnarray}
for the case of  $d$-wave pairing.

From Eqs. (\ref{gapswmf}), (\ref{gapdwmf}) it is easy to obtain also
appropriate equations for $T_c$. For the case of $s$-wave pairing we get:
\begin{equation}
1=\lambda\left\{\tilde\alpha\frac{2}{W^2}\int_{0}^{\infty}dDDe^{-\frac{D^2}
{W^2}}\int_{0}^{\omega_c}d\xi\frac{th\frac{\sqrt{\xi^2+D^2}}{2T_c}}
{\sqrt{\xi^2+D^2}}+(1-\tilde\alpha)\int_{0}^{\omega_c}d\xi\frac{th
\frac{\xi}{2T_c}}{\xi}
 \right\}
\label{tcequs}
\end{equation}
For the case of $d$-wave pairing in Eq. (\ref{tcequs}) we have only to
replace $\tilde\alpha$ by ``effective'' $\alpha_d$  from (\ref{alphd}). 
These equations for $T_c$ coincide with those obtained in microscopic
derivation of Ginzburg -- Landau expansion for this model in Ref.\cite{PS}, 
where these equations were studied in detail. In general we always have
$T_{c\infty}<T_c<T_{c0}$.

Temperature dependences of average gap $<\Delta>$  and mean -- field gap
$\Delta_{mf}$, obtained numerically from equations of our model for the case
of $s$-wave pairing, are shown in Fig. 2 \footnote{In case of $d$-wave
pairing temperature dependences of $<\Delta>$ and $\Delta_{mf}$ are
qualitatively similar to those obtained for $s$-wave pairing.}. 
Mean -- field gap $\Delta_{mf}$ goes to zero at $T=T_c<T_{c0}$, while
$<\Delta>$ is non zero up to $T=T_{c0}$, the ``tails'' in temperature
dependences of $<\Delta>$ in the region of $T_c<T<T_{c0}$ apparently
signifying the existence of local superconducting ``drops'' in the sample,
while superconductivity in a sample as a whole is absent. Note that
temperature dependences of $<\Delta(T)>$ shown in Fig. 2 qualitatively
resemble those observed in underdoped cuprates in ARPES \cite{RC,Ding} and
specific -- heat experiments \cite{Loram}, if we assume that real $T_c$
observed in these samples corresponds to our mean -- field $T_c$, while
``drops'' with $<\Delta>\neq 0$ exist for all temperatures $T>T_c$ up to
$T_{c0}$, which is significantly larger than $T_c$. This interpretation
implicitly assumes that in the ``absence'' of the pseudogap underdoped
cuprates would have possessed much larger temperature of superconducting
transition, than those observed in the experiment.

Despite the fact that superconductivity in a sample as a whole for
$T_c<T<T_{c0}$ (according to our interpretation) is absent, the presence of
non zero average gap $<\Delta>$ leads, as will be shown below, to the
appearance of a number anomalies in experimentally measurable 
characteristics, such as tunnelling density of states and spectral density
measured in ARPES -- experiments.

\section{Spectral density and density of states.}

Retarded Green's function of an electron in the vicinity of ``hot'' patch
on the Fermi surface in superconducting state is given by:
\begin{equation}
G^R(E,\xi_p)=\int_{0}^{\infty}dD{\cal P}(D)
\frac{E+\xi_p}{(E+i0)^2-\xi_p^2-D^2-\Delta^2(D)e^2(\phi)}
\label{Gr}
\end{equation}
From this we obtain spectral density as:
\begin{equation}
A(E,\xi_p)=-\frac{1}{\pi}ImG^R(E,\xi_p)=
\frac{2}{W^2}\int_{0}^{\infty}dDDe^{-\frac{D^2}{W^2}}(E+\xi_p)\delta(
\xi_p^2+D^2+\Delta^2(D)e^2(\phi)-E^2)
\label{spden}
\end{equation}
Using mean -- field approach, assuming $\Delta=\Delta_{mf}$ which is
independent of $D$, we obtain:
\begin{equation}
A_{mf}(E,\xi_p)=\frac{|E|+\xi_pSignE}{W^2}\exp\left(\frac{
\xi_p^2+\Delta_{mf}^2e^2(\phi)-E^2}{W^2}\right)\theta(E^2-\xi_p^2-
\Delta_{mf}^2e^2(\phi))
\label{spdenmf}
\end{equation}
In this approximation spectral density acquires a gap for
$|E|<\Delta_{mf}$ which disappears for $T\to T_c (\Delta_{mf}\to 0)$. 
In fact we have already seen, that $\Delta$ possesses a significant
dependence on the value of dielectric gap $D$ (cf.(\ref{gapsw}),(\ref{gapdw})), 
so that from (\ref{spden}) we get:  
\begin{equation}
A(E,\xi_p)=\sum_i\frac{|E|+\xi_p SignE}{W^2}e^{-\frac{D_i^2}{W^2}}
\frac{1}{\left| 1+\frac{d\Delta^2(D)}{dD^2}|_{D=D_i}e^2(\phi)\right| }
\label{spdenex}
\end{equation}
where $D_i$ -- are the positive roots of the equation
$D^2+\xi_p^2+\Delta^2(D)e^2(\phi)-E^2=0$. Energy dependences of spectral
density for $\xi_p=0$, i.e. for an electron momentum on the Fermi surface
(we restrict ourselves only to this case) are given in Fig. 3 and Fig. 4 for 
the case of $s$-wave and $d$-wave pairing respectively.

For $T_{c\infty}<T<T_{c0}$ spectral density acquires a discontinuity at
$E=D_{max}$ due to discontinuity of the derivative $d\Delta^2(D)/dD^2$ at
$D=D_{max}$ (i.e. at maximal $D$ for which $\Delta(D)$ is non zero). 
Effects of finite correlation length of fluctuations $\xi$ will
obviously lead to the smearing of this discontinuity, however there will
remain a characteristic dip after the main peak of the spectral density.
Similar dip is observed in ARPES -- experiments \cite{Tim,RC} and there is
no accepted interpretation of it up to now. 

In the case  $d$-wave pairing the value of $D^2+\Delta^2(D)$ grows with the
growth of $D$, thus the equation $D^2+\Delta^2(D)-E^2=0$ acquires roots only
for $|E|>\Delta_0$. Thus, the gap in the spectral density appears for
$|E|<\Delta_0$, so that the width of this gap is determined by $\Delta_0$, 
and not by $\Delta_{mf}$. Besides, the gap in the spectral density appears 
for $T=T_{c0}$, and there is no qualitative changes in the spectral density 
at $T=T_c$.

In case of $d$-wave pairing, for small enough width of the pseudogap
$W$ and small fraction of flat patches $\alpha_d$, the value of 
$D^2+\Delta^2(D)e^2(\phi)$ grows with the growth of $D$ and the width of the
gap in spectral density becomes equal to $\Delta_0e(\phi)$, analogously to 
the case of $s$-wave pairing. However, with the growth of the width of the
pseudogap $W$ and the fraction of flat patches, the value of 
$D^2+\Delta^2(D)e^2(\phi)$ drops with the growth of $D$ for small enough
$D$, leading to the width of the gap in spectral density smaller than 
$\Delta_0$, while for $E=\Delta_0$ there is a discontinuity in spectral
density (discontinuity at $E=D_{max}$ remains).

Consider now the tunnelling density of states $N(E)$. In case of $s$-wave
pairing we obtain:
\begin{eqnarray}
\frac{N(E)}{N_0(0)}=\frac{2}{W^2}\int_{0}^{\infty}dDDe^{-\frac{D^2}{W^2}}
\left\{\tilde\alpha\frac{|E|}{\sqrt{E^2-D^2-\Delta^2(D)}}\theta(E^2-D^2-
\Delta^2(D))+\right. \nonumber\\
\left.+(1-\tilde\alpha)\frac{|E|}{\sqrt{E^2-\Delta^2(D)}}\theta(E^2-
\Delta^2(D))\right\}
\label{dossw}
\end{eqnarray}
Assuming self -- averaging nature of the gap we have
$\Delta=\Delta_{mf}$ which does not depend on fluctuations of $D$, then:  
\begin{equation} 
\frac{N_{mf}(E)}{N_0(0)}=\left\{\tilde\alpha\frac{2}{W^2}
\int_{0}^{\sqrt{E^2-\Delta_{mf}^2}}dDDe^{-\frac{D^2}{W^2}}\frac{|E|}
{\sqrt{E^2-D^2-\Delta_{mf}^2}}+(1-\tilde\alpha)\frac{|E|}{\sqrt{E^2-
\Delta_{mf}^2}}\right\}\theta(E^2-\Delta_{mf}^2).
\label{dosswmf}
\end{equation}
In this approximation for $|E|<\Delta_{mf}$ there is a gap in the density
of states which vanishes for $T\to T_c(\Delta_{mf}\to 0)$, however there  
remains the pseudogap due to AFM -- fluctuations:
\begin{equation} 
\frac{N(E)}{N_0(0)}=\tilde\alpha\frac{2}{W^2}\int_{0}^{E}dDDe^{-\frac{D^2}
{W^2}}\frac{|E|}{\sqrt{E^2-D^2}}+(1-\tilde\alpha)
\label{dosafm}
\end{equation}
discussed previously in Ref. \cite{PS}. In fact $\Delta(D)$ in (\ref{dossw})
depends on $D$ according to (\ref{gapsw}). From Eq. (\ref{dossw}) and
appropriate dependence $\Delta(D)$ it can be seen that for $T<T_{c\infty}$ 
there is a gap in the density of states for $E<\Delta_{\infty}$, while for
$T>T_{c\infty}$ there is no gap in the density of states, but certain
contribution to the pseudogap due to superconducting pairing remains.
For $T_c<T<T_{c0}$ the gap $\Delta(D)$ is different from zero for
$D<D_{max}$, thus the difference from pseudogap behavior due only to
AFM -- fluctuations is observed already for $T_c<T<T_{c0}$, and only for
$T>T_{c0}$ we observe purely AFM -- pseudogap (\ref{dosafm}).

In Fig. 5 we show energy dependence of the density of states in case of
$s$-wave pairing for different temperatures. There is a cusp in the
density of states at $|E|=\Delta_0$, besides for $T>T_{c\infty}$ there is
another cusp at $|E|=D_{max}>\Delta_0$, though this last cusp is only
visible for high enough temperatures $T\sim T_{c0}$.  Density of states
changes qualitatively only at $T=T_{c0}$ while nothing special happens at
mean -- field $T_c$.

For $d$-wave pairing density of states becomes:
\begin{eqnarray}
\frac{N(E)}{N_0(0)}=\frac{4}{\pi}\frac{2}{W^2}\int_{0}^{\infty}dDD
e^{-\frac{D^2}{W^2}}\left\{\int_{0}^{\alpha}d\phi\frac{|E|}
{\sqrt{E^2-D^2-\Delta^2(D)e^2(\phi)}}\theta(E^2-\Delta^2(D)e^2(\phi)-D^2)+
\right.\nonumber\\ 
\left.+\int_{\alpha}^{\pi/4}d\phi\frac{|E|}{\sqrt{E^2-
\Delta^2(D)e^2(\phi)}}\theta(E^2-\Delta^2(D)e^2(\phi))\right\}
\label{dosdw}
\end{eqnarray}
Assuming self -- averaging $\Delta=\Delta_{mf}$ and is independent of $D$.
Then the width of superconducting pseudogap in the density states is of the
order of $\Delta_{mf}$ and the appropriate contribution vanishes for
$T\to T_c$ while the pseudogap (\ref{dosafm}) due to AFM -- fluctuations 
remains. However, in reality in (\ref{dosdw}) we have $\Delta=\Delta(D)$
defined by Eq. (\ref{gapdw}). 

The behavior of the density of states in case of $d$-wave pairing is shown
in Fig. 6. As for $s$-wave pairing we observe significant difference
between the exact density of states and that obtained in the mean -- field
approach, which is due to fluctuations of superconducting gap
(superconducting ``drops'') induced by AFM short -- range order.
Exact density of states does not ``feel'' superconducting transition in a 
sample as a whole which takes place at $T=T_c$. Characteristic width of the
pseudogap in the density states is given by $\Delta_0$, not by $\Delta_{mf}$,
as in mean -- field approximation. Appropriate contributions become
observable already at $T=T_{c0}>T_c$.

\section{Conclusion.}

In this work we have continued our studies of anomalies of superconducting
state in quite simplified exactly solvable model of the pseudogap in 
two -- dimensional system of electrons \cite{PS}. The main simplifying
assumption of our model (in addition to the static nature of fluctuations)
is the use of the limit of $\xi\to\infty$ for correlation length of
fluctuations of AFM short -- range order, which allows us to obtain main
results in analytic form. In particular, in this limit we can obtain an
exact average superconducting gap (\ref{avergap}).  Though it is clear,
in principle, that this model can be generalized to the case of finite
correlation lengths \cite{Sch,KS,C99}, it is less clear how we will be able 
to analyze superconducting state in this generalized model outside the
limits of mean -- field approach, similar to what was done above for 
the case of $\xi\to\infty$. Qualitatively it is clear that the effects of
finite $\xi$ will lead to some smearing of cusps and discontinuities which
appear in the model with $\xi\to\infty$, as well as relatively smooth
dependence of $T_c$ and other characteristics of superconducting state on
the value of $\xi$.

The results obtained above show that the pseudogap state induced by
AFM short -- range order fluctuations (or similar CDW fluctuations) leads
(in addition to the anomalies of the normal state \cite{Sch,KS,C99})
also to rather unusual properties of superconducting state, related to
partial dielectrization (non Fermi -- liquid behavior) of electronic 
spectrum on the ``hot'' patches of the Fermi surface. These properties
correlate well with a number of anomalies observed in the underdoped state
of HTSC -- cuprates. It is obvious that more serious comparison with
experiments can only be performed in more realistic approach, taking into
account, first of all, the effects of the finite correlation lengths
$\xi$, which is relatively small in real systems. At low temperatures it is
also important to take into account the dynamic nature of AFM fluctuations.

This work is supported in part by the grant of the Russian Foundation for
Basic Research No. 99-02-16285, as well as by the project No. 020 
of the State Programs ``Statistical Physics''  and project No. 96-051
of the State Program on HTSC of the Russian Ministry of Science. 

\newpage
\begin{center}
{\bf Figure Captions:}
\end{center}

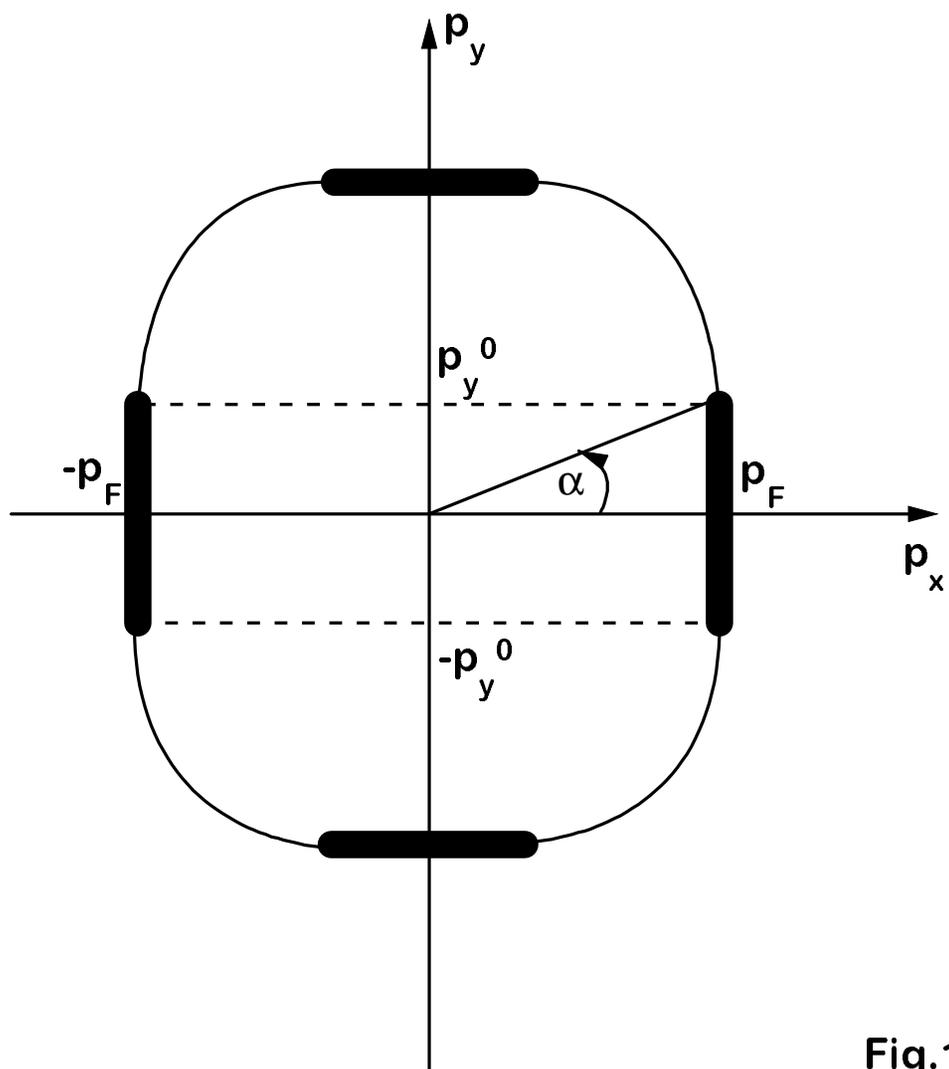
Fig.1. Model Fermi surface of two -- dimensional system. ``Hot'' patches are
shown by thick lines of the width of $\sim \xi^{-1}$.

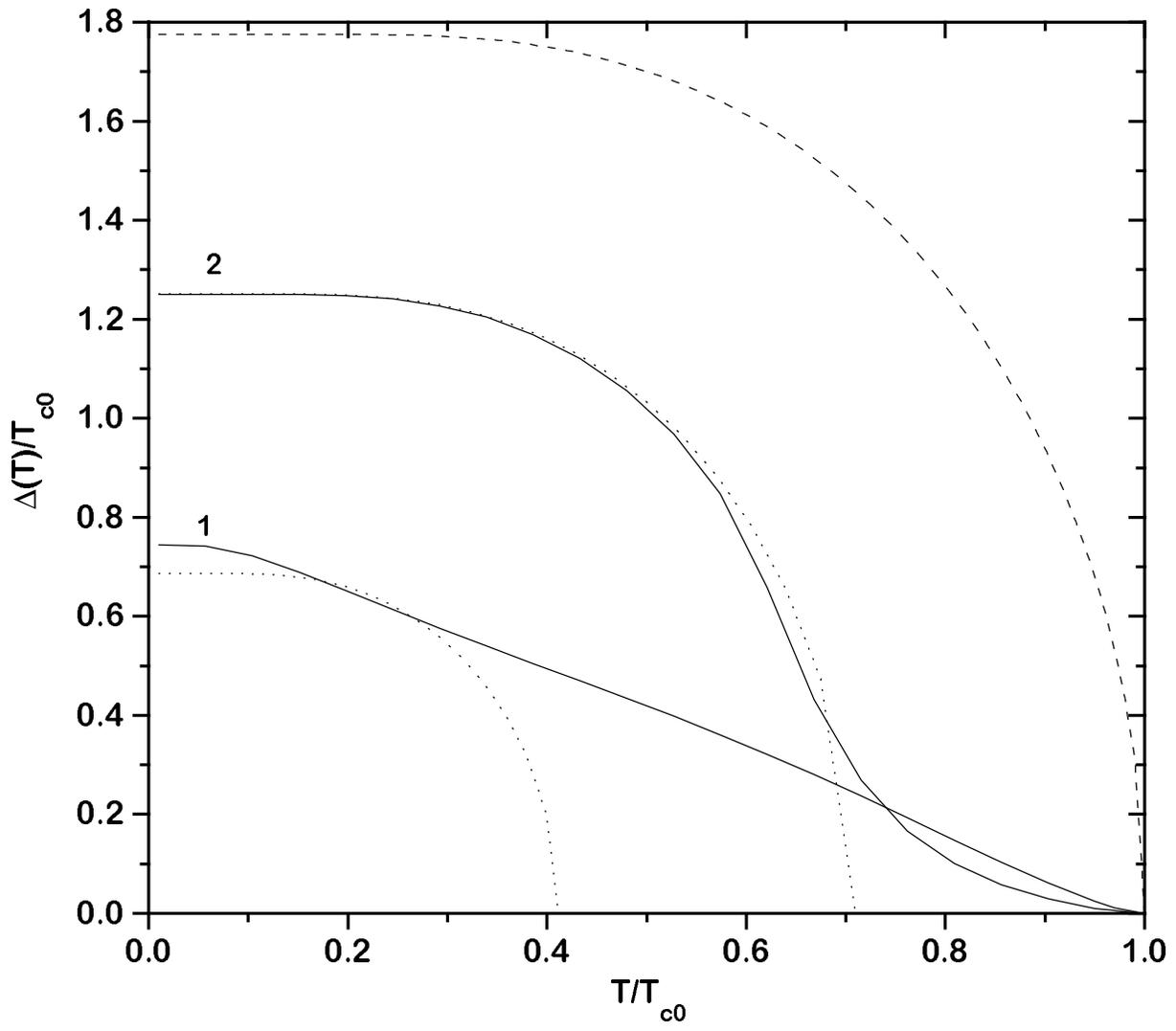
Fig.2. Temperature dependences of superconducting gaps  $\Delta_{mf}$ 
(points), $<\Delta>$ (full lines) and $\Delta_{0}$ (dashed line) in case of
$s$-wave pairing.\\
1.--- $\lambda =0,4$; $\tilde\alpha =2/3$; $\omega_{c}/W=3$ ($T_{c}/T_{c0}=0,42$).\\
2.--- $\lambda =0,4$; $\tilde\alpha =0,2$; $\omega_{c}/W=1$ ($T_{c}/T_{c0}=0,71$).

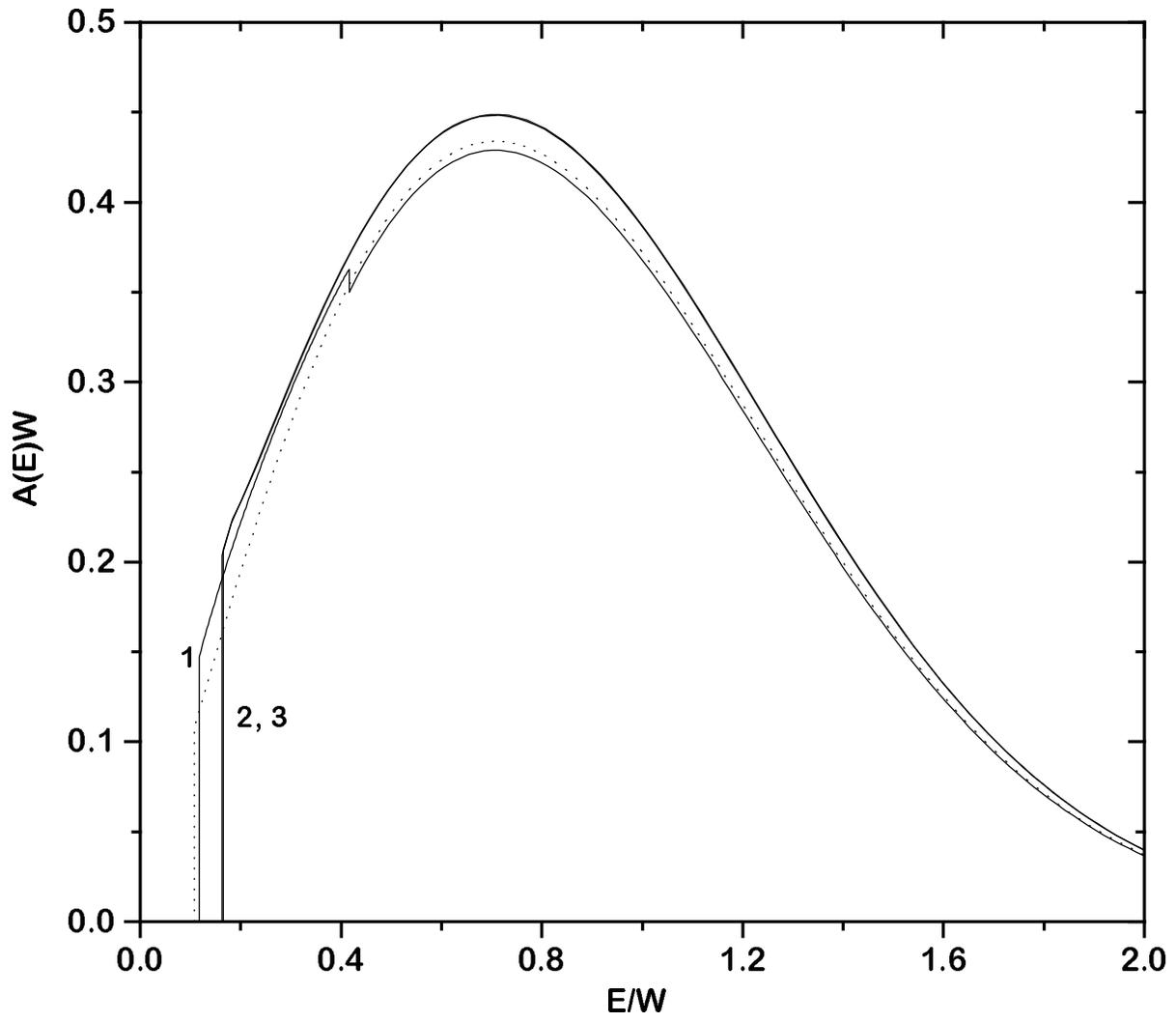
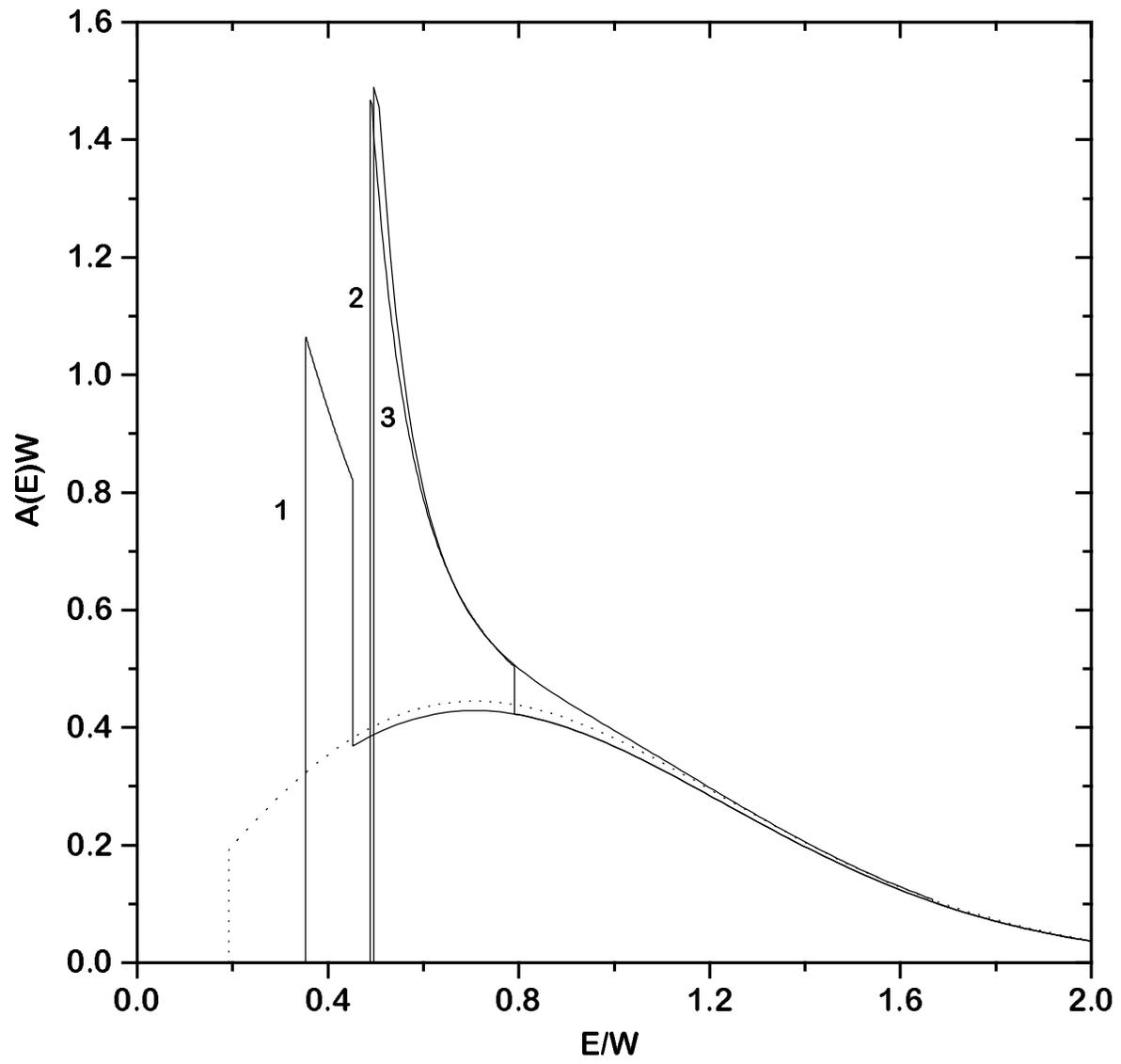
Fig.3. Spectral density on the Fermi surface in case of $s$-wave pairing for
different values of $T/T_{c0}$: 1.--0,8; 2.--0,4; 3.--0,1.

(a) --- $\lambda =0,4$; $\tilde\alpha =0,2$; $\omega_{c}/W=1$ 
($T_{c}/T_{c0}=0,71$, $T_{c\infty}/T_{c0}=0,54$).\\
Points: mean -- field approximation for the spectral density 
$A_{mf}(E)$ for $T/T_{c0}=0,4$.

(b) --- $\lambda =0,4$; $\tilde\alpha =2/3$; $\omega_{c}/W=3$ 
($T_{c}/T_{c0}=0,42$, $T_{c\infty}/T_{c0}=7\cdot 10^{-3}$).\\
Points: mean -- field approximation for $A_{mf}(E)$ at $T/T_{c0}=0,1$

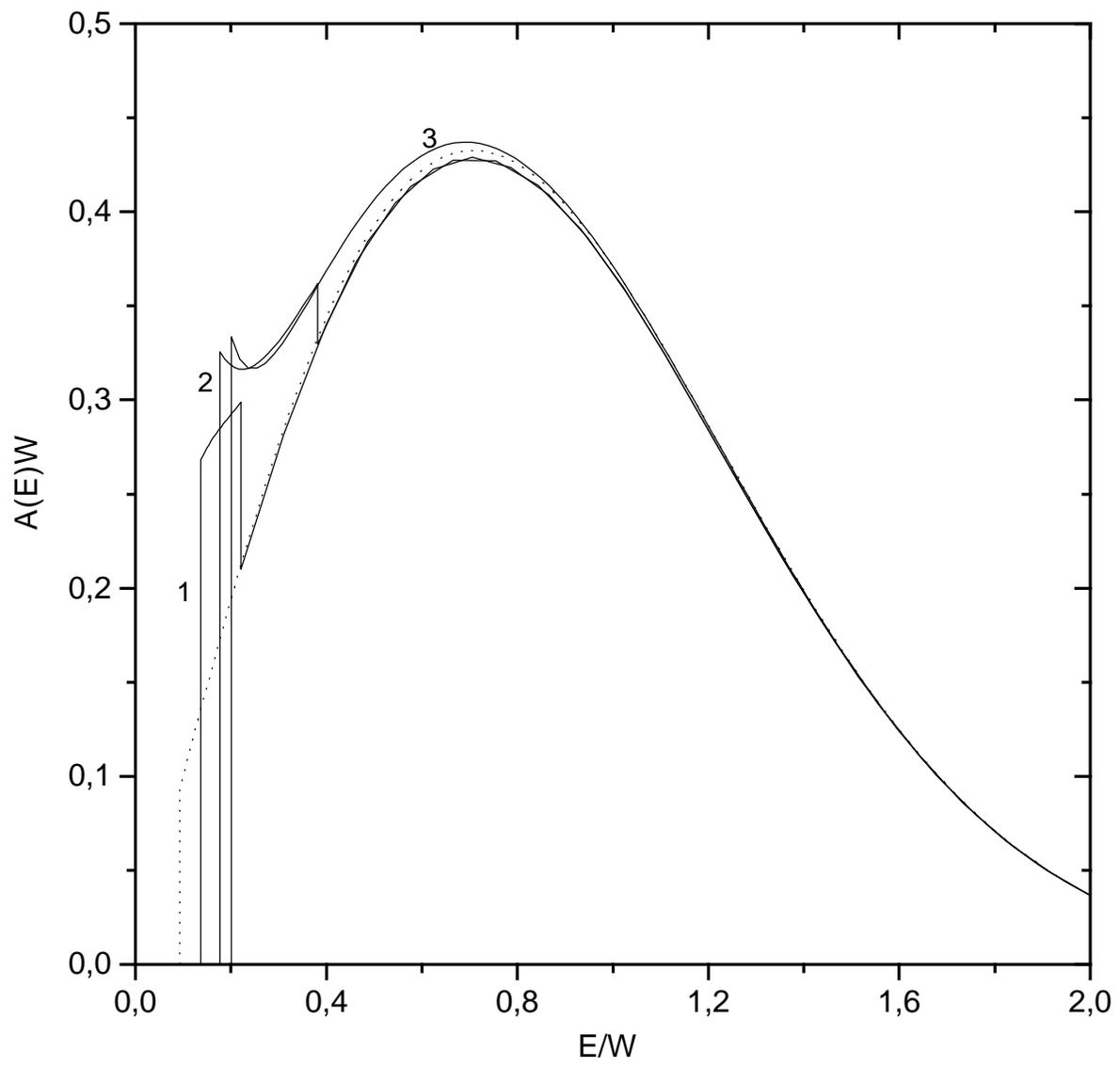
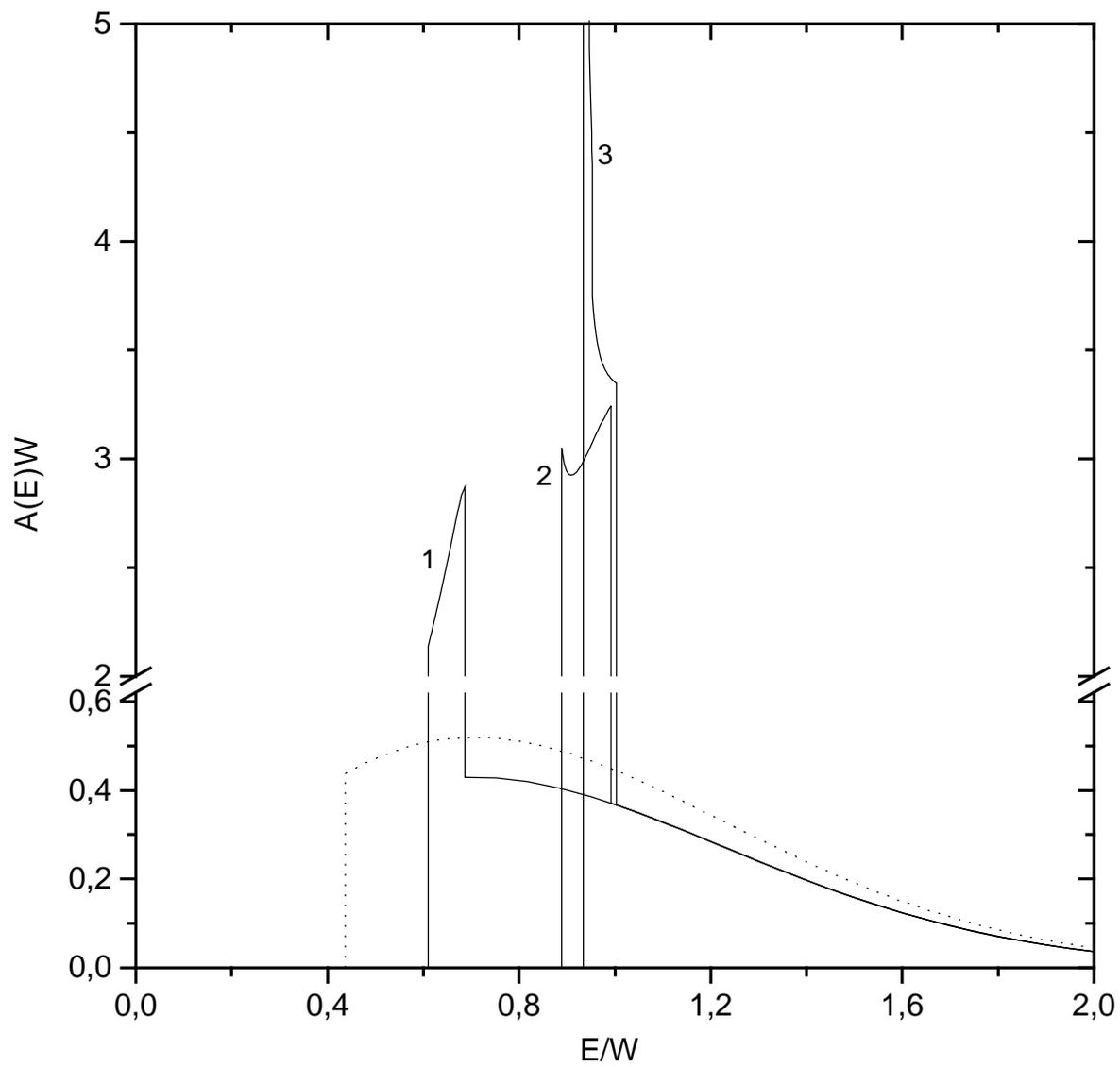
Fig.4. Spectral density on the Fermi surface for $\phi =0$ in case of 
$d$-wave pairing.

(a) --- $\lambda =0,4$; $\tilde\alpha =0,2$; $\omega_{c}/W=1$ 
($T_{c}/T_{c0}=0,42$, $T_{c\infty}/T_{c0}=0,2$),\\
$T/T_{c0}=$: 1.--0,8; 2.--0,6; 3.--0,1.

(b) --- $\lambda =0,4$; $\tilde\alpha =2/3$; $\omega_{c}/W=5$ 
($T_{c}/T_{c0}=0,48$, $T_{c\infty}/T_{c0}\sim 10^{-18}$),\\
$T/T_{c0}=$: 1.--0,8; 2.--0,3; 3.--0,1.
Points: mean -- field spectral density $A_{mf}(E)$ for $T/T_{c0}=0,1$

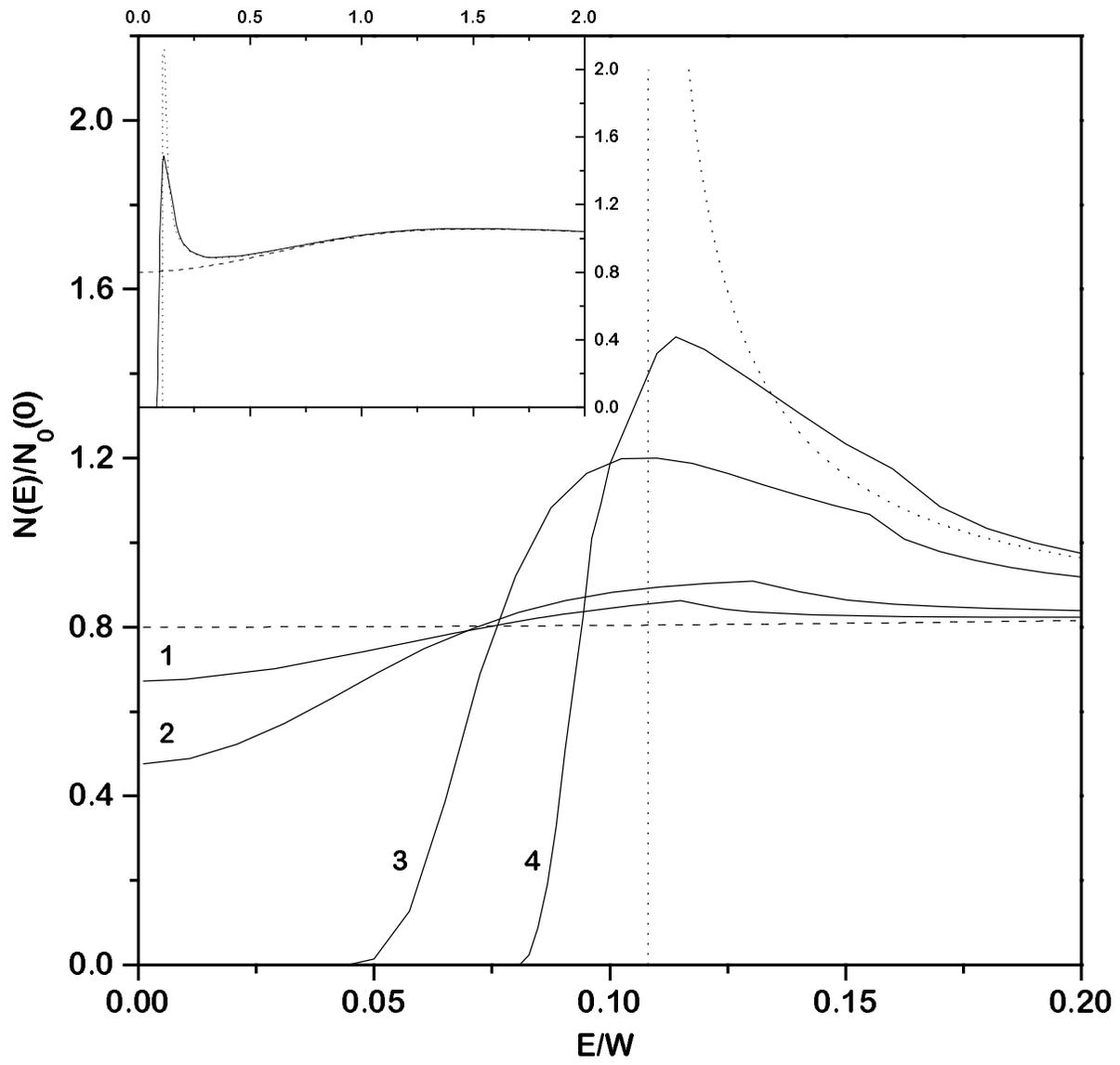
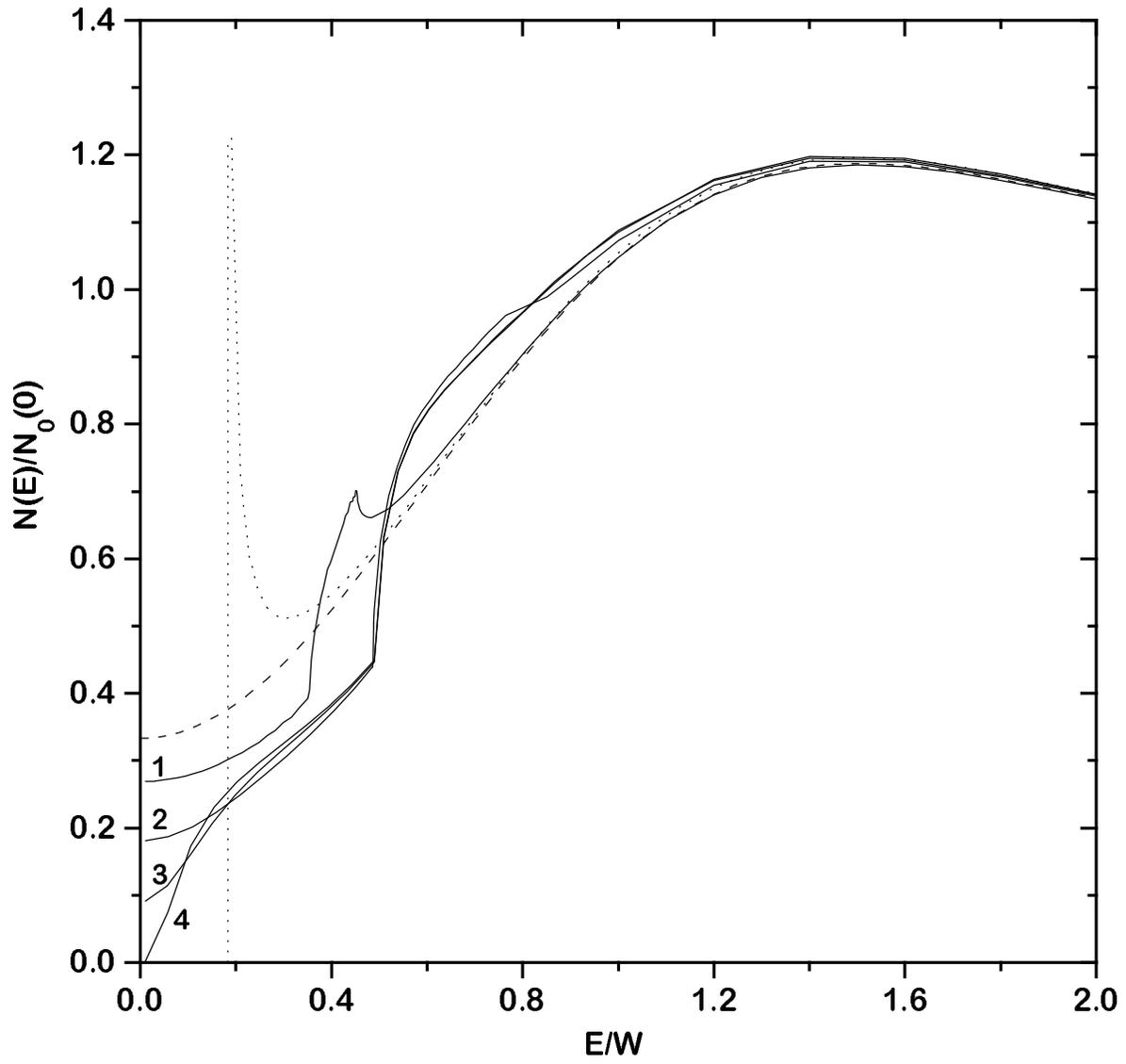
Fig.5. Density of states in case of $s$-wave pairing.

(a) --- $\lambda =0,4$; $\tilde\alpha =0,2$; $\omega_{c}/W=1$ 
($T_{c}/T_{c0}=0,71$, $T_{c\infty}/T_{c0}=0,54$),\\
$T/T_{c0}=$: 1.--0,8; 2.--0,71; 3.--0,54; 4.--0,1.\\
Points: mean -- field density of states $N_{mf}(E)$ for $T/T_{c0}=0,4$\\
At the insert: density of states for $T/T_{c0}=0,4$

(b) --- $\lambda =0,4$; $\tilde\alpha =2/3$; $\omega_{c}/W=3$ 
($T_{c}/T_{c0}=0,42$, $T_{c\infty}/T_{c0}=7\cdot 10^{-3}$),\\
$T/T_{c0}=$: 1.--0,8; 2.--0,42; 3.--0,2; 4.--0,05.\\
Points: mean field density of states $N_{mf}(E)$ for $T/T_{c0}=0,1$.
Dashed line: pseudogap in the density of states for $T>T_{c0}$.

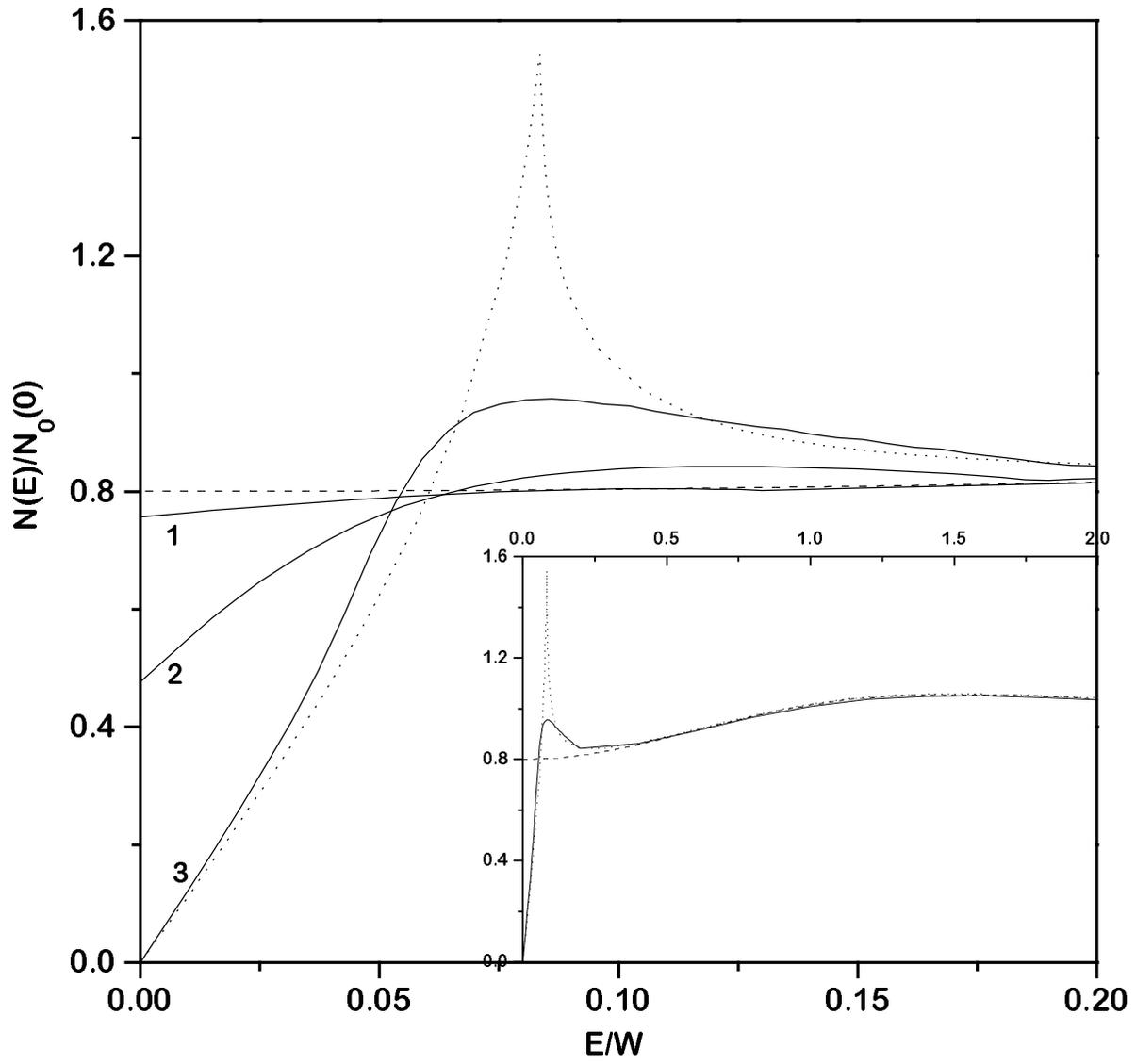
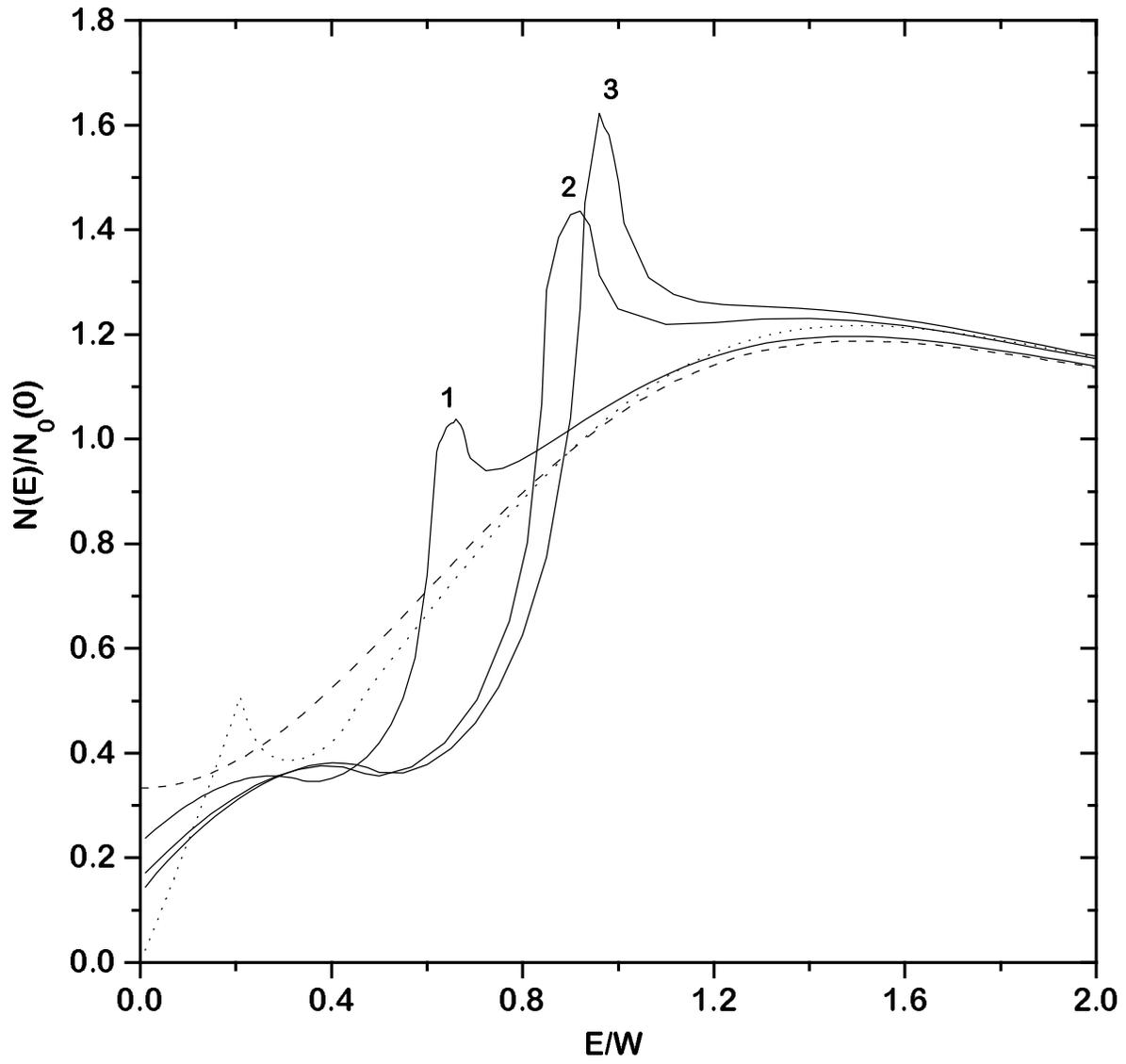
Fig.6. Density of states in case of $d$-wave pairing.

(a) --- $\lambda =0,4$; $\tilde\alpha =0,2$; $\omega_{c}/W=1$ 
($T_{c}/T_{c0}=0,42$, $T_{c\infty}/T_{c0}=0,2$),\\
$T/T_{c0}=$: 1.--0,8; 2.--0,42; 3.--0,2.\\
Points: mean field density of states $N_{mf}(E)$ for $T/T_{c0}=0,2$\\
At the insert: density of states for $T/T_{c0}=0,2$.

(b) --- $\lambda =0,4$; $\tilde\alpha =2/3$; $\omega_{c}/W=5$ 
($T_{c}/T_{c0}=0,48$, $T_{c\infty}/T_{c0}\sim 10^{-18}$),\\
$T/T_{c0}=$: 1.--0,8; 2.--0,48; 3.--0,1.\\
Points: mean field density of states $N_{mf}(E)$ for $T/T_{c0}=0,1$.
Dashed line: pseudogap in the density of states for $T>T_{c0}$.

\end{document}